\documentclass[11pt,twoside]{article}
\usepackage{asp2010}

\resetcounters

\begin{document}

\title{VAST - a real-time pipeline for detecting radio transients and variables on the Australian SKA Pathfinder (ASKAP) telescope}
\author{Jay~Banyer$^{1,2}$, Tara~Murphy$^{1,2,3}$, and~the~VAST~Collaboration}
\affil{$^1$Sydney Institute for Astronomy (SIfA), School of Physics, The University of Sydney, NSW 2006, Australia}
\affil{$^2$ARC Centre of Excellence for All-sky Astrophysics (CAASTRO), The University of Sydney, NSW 2006, Australia}
\affil{$^3$School of Information Technologies, The University of Sydney, NSW 2006 Australia}

\begin{abstract}
The Australian SKA Pathfinder (ASKAP) is a next generation radio telescope currently under construction in Western Australia. The fast survey speed and wide field of view make it an ideal instrument for blind transients searches. The ASKAP Variable and Slow Transients (VAST) survey is a one of the major science programs planned for ASKAP. The scientific goals of VAST include the detection and characterisation of a wide range of transient and variable phenomena, from gamma-ray burst afterglows to extreme scattering events, on timescales of 5 seconds or longer.

We describe the data and processing challenges involved in running the VAST real-time transient detection pipeline. ASKAP will produce 2.5 GB of visibility data per second, transformed into one 8GB image cube every 5 seconds. Each cube will contain approximately twenty 100 megapixel images with 100s of radio sources detected in each epoch. The VAST pipeline will measure and monitor all of these sources, detect variables and transients and generate alerts using the VOEvent framework.

The goal of the VAST Design Study is to develop a prototype pipeline to establish and demonstrate the functionality of the final ASKAP pipeline. We give an overview of the prototype pipeline's functionality, technical implementation and current status.
\end{abstract}

\section{The Australian SKA Pathfinder (ASKAP)}
The Australian SKA Pathfinder \citep[ASKAP, ][]{johnston2008} is a powerful radio telescope currently under construction in Western Australia. It is a 36 dish radio interferometer with exceptional survey speed: it can image the entire southern sky to high sensitivity in two nights. The fast survey speed and wide field of view make it an ideal instrument for blind transients searches.

ASKAP is located at the Murchison Radio-astronomy Observatory (MRO), approximately 315km north east of Geraldton, Western Australia. The MRO is also the location of the Murchison Widefield Array (MWA) telescope and is the candidate site for Australia and New Zealand's bid to host the Square Kilometre Array (SKA).

A 6 antenna array called BETA is currently being installed and will be commissioned during 2012. ASKAP is scheduled to begin scientific observations with a 6 to 12 antenna array during 2013.

\section{The VAST Survey}
ASKAP will give us an unprecedented opportunity to investigate the transient sky at radio wavelengths. Its wide-field survey capabilities will enable the discovery and investigation of variable and transient phenomena from the local to the cosmological, including flare stars, intermittent pulsars, X-ray binaries, magnetars, extreme scattering events, intra-day variables, radio supernovae and orphan afterglows of gamma ray bursts. In addition ASKAP probes unexplored regions of phase space where new classes of transient sources may be detected.

The VAST survey \citep{vastmemo1} is one of ten major science programs planned for ASKAP. It will involve several survey regimes including one that images most of the visible sky every night for two years. This will produce measurements of one million sources over hundreds of epochs.

\section{Transient Pipeline Functionality}
\begin{figure}[!ht]
\plotone{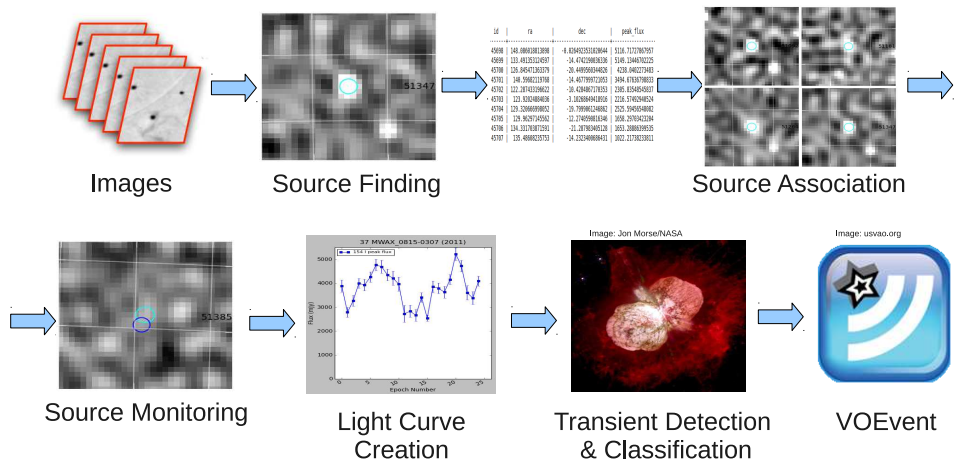}
\caption{Illustration of VAST pipeline functionality\label{fig:pipeline}}
\end{figure}

The VAST transient pipeline will detect transients and variables by analysing the images produced by the ASKAP imaging pipeline. The VAST pipeline operates by extracting and measuring sources in images, constructing and analysing light curves and generating notifications of detections. This approach is based on techniques described by \citet{bannister2011}. The functionality of the VAST pipeline is illustrated in Figure~\ref{fig:pipeline} and described in Table~\ref{tab:functionality}.

The VAST collaboration is also investigating the techniques of transient detection using image subtraction and visibility domain transient detection \citep{trott2011}.

\begin{table}[!ht]
\caption{Description of VAST pipeline functionality\label{tab:functionality}}
\begin{center}
\begin{tabular}{p{1.5in}p{3.5in}}
\tableline
\noalign{\smallskip}
Function & Description\\
\noalign{\smallskip}
\tableline
\noalign{\smallskip}
Source finding &
Blind source finding is performed on every image. This produces a list of detected sources with position, flux (brightness) and size.\\
\noalign{\smallskip}
Source association &
Each detected source is associated with one from the master list by finding the master source at the same position as the detection. If no master source is found at that position then a new master source is added.\\
\noalign{\smallskip}
Image quality control &
The image is assessed for quality by comparing the measured parameters of the detected sources to a list of well-characterised sources using a technique described by \citet{bannister2011}.\\
\noalign{\smallskip}
Source monitoring &
Master sources known to be in the field that were not detected by the source finder are measured by attempting to fit a Gaussian at the position where the source is expected to be. If no source can be fit then an upper-limit on the flux at that position is taken instead.\\
\noalign{\smallskip}
Light curve creation &
All measurements for each source are collated into a radio light curve - a collection of flux measurements for all epochs where the source was observed.\\
\noalign{\smallskip}
Detect and classify transients and variables &
The updated light curves are scanned to detect transient or variable behaviour. Sources that exhibit transient or variable behaviour are classified using a machine-learning algorithm. This attempts to determine what kind of physical phenomenon is causing the variability.\\
\noalign{\smallskip}
Notification &
Detections and classifications are subjected to quality control and cross-matching to external catalogues. Those that appear genuine are added to an archive and the community is alerted using VOEvent notifications.\\
\noalign{\smallskip}
\tableline
\end{tabular}
\end{center}
\end{table}

\section{Capacity Challenges}
The traditional model for radio astronomy is to take the data from the telescope and process it off-site some time later. ASKAP produces too much data to store and transport so the VAST transient survey will be implemented as a near real-time pipeline running on the same large computing cluster as the ASKAP imaging systems.

ASKAP's imaging pipeline will produce two streams of images: one with integration times of 5 seconds, the other with longer integration times depending on the observing mode e.g. 1 minute, 1 hour, 8 hours. Each 5 second image cube will contain approximately twenty \textasciitilde100 megapixel images: \textasciitilde5 frequency bands each with 4 polarisations. Each cube will be \textasciitilde8GB in size.

In the transient pipeline the 5 second image stream will require approximately 12,000 source measurements per second (approximately 100 sources per square degree, an image size of 30 square degrees, 20 images per cube, one cube every 5 seconds). Light curves for approximately 3,000 sources will change each second. Each source detection requires a cone search to associate the detection with a known source position. 3,000 cone searches will be performed each second however most of these can be optimised by searching a cached list of the known sources in the current field. The cached list will be refreshed each time the telescope moves to a new field, which will happen no more frequently than every minute or so.

The image stream with longer integrations (e.g. 1 hour) will produce images at a lower rate but each image will contain more sources due to higher sensitivity. The source measurement load for this stream will vary depending on the integration time.

\section{VAST Pipeline Prototype}
A prototype transient pipeline has been developed as part of the VAST Design Study. It allows iterative development of the functional requirements of the final ASKAP pipeline. The prototype pipeline will also be used to perform transient detection on data from other telescopes, including: the Murchison Widefield Array (MWA), the Australia Telescope Compact Array (ATCA), the Very Large Array (VLA), and the SKA Molonglo Prototype (SKAMP).

The prototype takes a series of FITS images as input, detects and measures sources, performs image quality control, constructs light curves and outputs results to a database. Transient detection, classification and VO alerts are being developed.

The prototype uses a custom source finder that is able to characterise multiple overlapping sources more accurately than other existing source finders (Hancock et al. in prep).

A dynamic website provides a user interface to the database. Its purpose is to help visualise and diagnose the pipeline output easily and efficiently. It can browse the lists of images, master sources and detections, plot light curves, create postage stamps and perform cross-matching with external catalogues. 

The prototype has already proven to be a useful tool for analysing datasets from radio telescopes even when not looking for transients due to its ability to measure and report on all sources in all images with very little effort. This assists with checking the image quality, calibration and positional accuracy of the dataset.

\subsection{Languages, Tools and Libraries}
The prototype pipeline is developed using the Python programming language and the Django web framework. These were chosen for their suitability for rapidly changing requirements during the VAST Design Study and the availability of astronomical libraries. PostgreSQL is used for the SQL database with the Q3C plugin for optimised celestial coordinate searches (cone searches).

\bibliography{O04}

\end{document}